\documentclass[ aps, pra, amsmath,amssymb, 10pt,  final, tightenlines,  twoside,  twocolumn,  nofloats, nofootinbib,  superscriptaddress, showkeys,
showkeywords]{revtex4-2}

\usepackage[normalem]{ulem}
\usepackage{amssymb,amsfonts,amsmath,amstext,graphicx,lipsum,empheq}
\usepackage{float}
\usepackage[normalem]{ulem}

\usepackage{outlines}
\usepackage{physics}
\usepackage[dvipsnames]{xcolor}
\usepackage[colorlinks=true,
            linkcolor=blue,
            citecolor=blue,
            filecolor=magenta,
            urlcolor=magenta]{hyperref}

\usepackage{subcaption}
\usepackage{ragged2e}
\usepackage{bm}

\usepackage{upgreek}
\newcommand{\tlife}{\uptau_s}
\newcommand{\tM}{\uptau_M}
\newcommand{\tc}{\uptau_c}



\begin{document}
\title{Experimentally relevant witness for non-Markovianity in discrete time crystals}
\date{\today}
\author{Rahul Ghosh}
\affiliation{Department of Chemistry, Ben Gurion University of the Negev, 1 Ben-Gurion Ave, Beer Sheva, 8410501, Israel}
\author{Victor Mukherjee}%
\affiliation{Department of Physical Sciences, Indian Institute of Science Education and Research Berhampur, Berhampur 760010, India}
\author{Yonatan Dubi}%
\affiliation{Department of Chemistry, Ben Gurion University of the Negev, 1 Ben-Gurion Ave, Beer Sheva, 8410501, Israel}
\affiliation{The Ilze-Katz Institute for Nano-Scale Science and Technology, Ben Gurion University of the Negev, 1 Ben-Gurion Ave, Beer Sheva, 8410501, Israel}

\begin{abstract}
Discrete time crystals (DTCs) are almost always analyzed under the Born-Markov approximation, in which the environment is memoryless. We show that a reservoir with a finite correlation time $\tau_m$ leaves an exact and strikingly simple imprint on the stroboscopic dynamics of DTCs, which can be used as an experimentally relevant witness for non-Markovianity. 
By analyzing a post-Markovian master equation we prove that the Floquet spectrum splits into the ordinary Markovian spectrum plus a replica of the purely unitary Floquet spectrum, uniformly contracted by $e^{-T/\tau_m}$, where $T$ is the drive period. 
Consequently, the DTC signal decay rate becomes independent of any system parameter once $\tau_m$ exceeds the Markovian lifetime of the DTC. The measured decay, and in particular its insensitivity to tunable system parameters, thus becomes a direct read-out of the reservoir correlation time. 
Below this threshold the dynamics is already non-Markovian, but the memory remains hidden in the decay. 
We confirm this in two structurally unrelated DTCs, a collective long-range model with incoherent pumping and a disorder-free Ising chain with dephasing, where a single estimator recovers $\tau_m$. The time crystal thereby witnesses the non-Markovianity of its own environment.
\end{abstract}
\maketitle

\section{Introduction}
The spontaneous breaking of time-translation symmetry gives rise to time crystals, phases of matter in which a many-body system develops persistent temporal order~\cite{wilczek2012_quantum,sacha2018_time}. While no-go theorems preclude such order in thermal equilibrium~\cite{bruno2013_impossibility,watanabe2015_absence}, periodically driven (Floquet) systems evade these constraints and host discrete time crystals (DTCs), in which the response locks to a subharmonic of the drive and the discrete time-translation symmetry is spontaneously broken down to a subgroup~\cite{else2016_floquet,khemani2016_phase,vonkeyserlingk2016_absolute,yao2017_discrete}. Following these early proposals, DTC order has been studied across a wide range of platforms~\cite{else2020_discrete,khemani2019_brief,zaletel2023_colloquium,sacha2015_modeling} and observed experimentally with trapped ions, nitrogen-vacancy centers, and superconducting processors~\cite{zhang2017_observation, choi2017_observation,mi2022_time,frey2022_realization}. 

Indeed, real experimental systems are never perfectly isolated, and understanding time-crystalline order in the presence of an environment has become a central theme. Counterintuitively, dissipation is not necessarily detrimental: engineered losses can themselves stabilize nonequilibrium order, as in boundary time crystals~\cite{iemini2018_boundary} and dissipative discrete time crystals~\cite{gong2018_discrete,gambetta2019_discrete}, as well as continuous time crystals that break a continuous time-translation symmetry~\cite{buca2019_non,carollo2020_nonequilibrium_many,cabot2024_continuous}. Such dissipative time crystals have recently been realized in driven atom-cavity and related setups~\cite{kessler2021_observation,kongkhambut2022_observation}. Dissipation nonetheless competes with the interactions that sustain temporal order, and sufficiently strong coupling to the environment eventually ``melts'' the time crystal, driving the system into a stationary or synchronized state with no subharmonic response.

Almost all studies of dissipative time crystals adopt the Markovian (Born--Markov) approximation, in which the environment is memoryless and the dynamics is generated by a time-local master equation. Physical reservoirs, however, possess a finite correlation time, and the resulting non-Markovian dynamics is characterized by a backflow of information from the environment into the system~\cite{breuer2009_measure,rivas2014_quantum_non,breuer2016_colloquium,devega2017_dynamics}.  Previous studies such as the time-dependent dissipation in mean-field limit~\cite{das2024_discrete} and with strong symmetry~\cite{ghosh2026_freezing} have suggested the beneficial effects of non-Markovian dynamics  for preserving DTCs in the presence of strong dissipation . However, whether such memory effects help or hinder nonequilibrium order  in more general systems remains largely unexplored, and in particular it is an open question whether environmental memory can counteract the dissipative melting of a time crystal.

In this Letter we show that environmental memory does not merely modify time-crystalline order. Rather, it ``writes" its own correlation time into the observable response. We model the open dynamics with a post-Markovian master equation carrying an exponential memory kernel of tunable memory time $\tau_m$ and embed it exactly into a time-local evolution on an enlarged space. The embedded generator is block-triangular, and we prove that its Floquet spectrum is the ordinary Markovian spectrum together with a replica of the purely \emph{unitary} Floquet spectrum contracted by $e^{-T/\tau_m}$, where $T$ is the period of the drive. Whenever $\tau_m$ exceeds the Markovian lifetime of the crystal, the replica sets the subharmonic decay and the measured signal  envelope becomes a direct, calibration-free read-out of the reservoir correlation time, rigid against the drive, the interactions, the dissipation rate and the system size, while the subharmonic phase stays locked at $\pi$. Below this threshold the memory is hidden: the dynamics is non-Markovian, yet the crystal decays as if the bath were memoryless. We verify this in two structurally unrelated DTCs, a collective Lipkin-Meshkov-Glick (LMG) model \cite{Lipkin1965,Meshkov1965,Glick1965} with incoherent pumping,  and a disorder-free short-range tilted Ising chain with dephasing, and show that a single estimator recovers $\tau_m$ in both.

\section{Theoretical framework}
\subsection{Memory-embedded dynamics} 
We take the reduced dynamics to obey a periodically driven extension of the Shabani--Lidar post-Markovian master equation
(PMME)~\cite{shabani2005_completely,budini2014_post,sutherland2018_non,
zhang2022_predicting,nila2026_continuous}, which incorporates a finite reservoir-correlation time while retaining a
time-local Markovian limit,
\begin{equation}
\dot{\rho}(t)=\mathcal L_H(t)[\rho(t)]
+\mathcal L_D\!\left[
\int_0^t dt'\,k(t-t')\,
\Phi(t,t')\rho(t')
\right],
\label{eq:pmme}
\end{equation}
with $\mathcal L_H(t)[\rho]=-i[H(t),\rho]$, dissipator
$\mathcal L_D[\rho]=\sum_i \left( L_i\rho L_i^\dagger
-\frac12\{L_i^\dagger L_i,\rho\} \right)$, 
Liouvillian
$\mathcal L(t)=\mathcal L_H(t)+\mathcal L_D$ and $\Phi(t,t')=
\mathcal T\exp\!\left[\int_{t'}^t ds\,\mathcal L(s)\right]$. The single-exponential kernel is $k(t)=\tau_m^{-1}e^{-t/\tau_m}$ setting the reservoir correlation time
$\tau_m$~\cite{maniscalco2006_non,maniscalco2007_complete}. As
$\tau_m\!\to\!0$, $k(t)$ becomes a delta function and Eq.~\eqref{eq:pmme}
reduces to the Lindblad equation $\dot\rho=\mathcal L(t)\rho$, so $\tau_m$
tunes continuously from memoryless to non-Markovian dynamics. The PMME can exhibit CP-indivisible dynamics, as demonstrated
for specific kernels and parameter regimes in
Ref.~\cite{sutherland2018_non}. For exponential-kernel PMME
dynamics, noncommutation between the Hamiltonian and
dissipative superoperators can lead to information
backflow~\cite{budini2014_post,zhang2022_predicting}.
Both models considered below feature such noncommutation.

The single pole
lets one trade the memory integral for a single auxiliary matrix $\eta$
(defined as the convolution itself,
$\eta(0)=0$) (for details see \cite{supplemental_mine2026}), in the spirit of Lindblad rate equations~\cite{budini2014_post}: the pair
$\mathfrak P=(\rho,\eta)^{\mathsf T}$ obeys the exact \emph{time-local} equation
\begin{equation}
\dot{\mathfrak P}(t)=\mathfrak L(t)\,\mathfrak P(t),\quad
\mathfrak L(t)=\begin{pmatrix}
\mathcal L_H(t) & \mathcal L_D\\[3pt]
\tau_m^{-1}\mathbb I & \mathcal L(t)-\tau_m^{-1}\mathbb I
\end{pmatrix},
\label{eq:embed}
\end{equation}
where $\mathfrak L$ is not a Lindbladian on a doubled space ($2d^2$). The auxiliary matrix is unnormalized, so it is not a density matrix. Because our system is periodic,
$H(t{+}T)=H(t)$, the dynamics is governed by the one-period propagator
$\mathfrak U_F(\tau_m)=\mathcal T e^{\int_0^T\mathfrak L(t)\,dt}$. Starting from the
empty-memory state $\mathfrak P(0)=(\rho_0,0)^{\mathsf T}$ it advances the stroboscopic
state as $\mathfrak P(nT)=\mathfrak U_F^{\,n}\,\mathfrak P(0)$, where $n=t/T$ denotes the stroboscopic period number~\cite{ghosh2025_quantum}. The lifetime of an observable is then set by the
slowest expressed Floquet
multiplier $\mu_s$ of $\mathfrak U_F(\tau_m)$,
$\tlife=-T/\ln|\mu_s|$ (in units of time; $-1/\ln|\mu_s|$ in driving periods) (see SM~\cite{supplemental_mine2026}). 
In the following we analyze the internal structure of the Floquet propagator
$\mathfrak U_F(\tau_m)$.

\subsection{Replica theorem}
The embedding, Eq.~\eqref{eq:embed}, has an exact structural consequence. In the new variable $\varsigma\equiv\rho-\eta$, we have
$\dot\varsigma(t)=\big(\mathcal L_H(t)-\tau_m^{-1}\big)\varsigma(t)$,
$\dot\eta(t)=\tau_m^{-1}\varsigma(t)+\mathcal L(t)\eta(t)$. Thus, $\varsigma$ undergoes unitary evolution multiplied by
a scalar decay at rate $\tau_m^{-1}$. The one-period
propagator in the $(\varsigma,\eta)$ representation is block
lower triangular, with $d^2$-dimensional diagonal blocks
$e^{-T/\tau_m}\mathsf U_F^{\mathrm u}$ and $\mathsf U_F^M$.
Here
$\mathsf U_F^{\mathrm u}=
\mathcal T e^{\int_0^T\mathcal L_H(t)\,dt}$
is the unitary Floquet channel, and
$\mathsf U_F^M=
\mathcal T e^{\int_0^T\mathcal L(t)\,dt}$
is the Markovian Floquet channel. Block triangularity implies that its spectrum is exactly
the union of the spectra of these diagonal blocks:
\begin{align}
\mathrm{spec}\,\mathfrak U_F(\tau_m)=
\mathrm{spec}\,\mathsf U_F^{M}\ \cup\
e^{-T/\tau_m}\,\mathrm{spec}\,\mathsf U_F^{\mathrm u},
\label{eq:replica}
\end{align}
Since $\eta(0)=0$ gives $\varsigma(0)=\rho_0$, the measurable state carries a unitarily evolving \emph{replica} of its own initial condition,
\begin{equation}
\rho(nT)=e^{-nT/\tau_m}\,(\mathsf U_F^{\mathrm u} )^n[\rho_0]+\eta(nT).
\label{eq:replica_state}
\end{equation}
Physically, this exponential-kernel construction admits an
exact representation in which dissipation switches on once,
at a random time with probability density
$k(t)=\tau_m^{-1}e^{-t/\tau_m}$, consistent with
Ref.~\cite{budini2014_post}.
The replica in Eq.~\eqref{eq:replica_state} is the
not-yet-switched contribution: it evolves unitarily with
weight $e^{-t/\tau_m}$, which depends only on $t$ and $\tau_m$.
Complete positivity follows from the Lindblad dilation
described in Ref.~\cite{supplemental_mine2026}, where an
auxiliary gate switches dissipation on irreversibly at
rate $\tau_m^{-1}$.

\subsection{Rigidity and information partition}
Equation~\eqref{eq:replica} has two physical consequences.
First, because $\mathsf U_F^{\mathrm u}$ is unitary, every
replica Floquet multiplier has modulus exactly $e^{-T/\tau_m}$.
When an expressed replica mode controls the asymptotic
subharmonic decay, the lifetime-setting Floquet multiplier
locks to $|\mu_s|=e^{-T/\tau_m}$.
The asymptotic decay rate of the subharmonic envelope
then measures the kernel memory time $\tau_m$: it is independent of the
drive amplitude, the interactions, the dissipation rates and
the system size at fixed $\tau_m$.
Second, each replica multiplier has the form
$\mu=e^{-T/\tau_m}e^{i\phi}$, with $e^{i\phi}$ an eigenvalue
of $\mathsf U_F^{\mathrm u}$: $\tau_m$ fixes its radius,
while the coherent dynamics alone fixes its phase.
Decay and oscillation are thus carried by
distinct spectral data:
the radius encodes the kernel memory time,
while the phase encodes the
oscillation frequency, with $\phi_{\rm DTC}=\pi$
for a period-doubled response.
The replica theorem alone does not establish robust phase locking.
Only the rate is universal: the amplitude of the replica
contribution depends on the initial state and on the measured
observable as well as on the system dynamics.

\subsection{Branch crossing and three regimes}
Let $\mu_M$ denote the slowest expressed oscillating Floquet
multiplier of the purely Markovian Floquet operator
$\mathsf U_F^M$, and $\tM\equiv-T/\ln|\mu_M|$ the corresponding
Markovian lifetime.
Because the two branches in Eq.~\eqref{eq:replica} compete
for the largest modulus,
provided this Markovian mode and at least one relevant
oscillating replica mode carry nonzero weight in the full
embedded response, and no slower oscillatory mode becomes expressed,
the modulus of the selected Floquet multiplier is
\begin{equation}
|\mu_s(\tau_m)|=\max\big\{\,|\mu_M|\,,\;e^{-T/\tau_m}\big\},
\label{eq:maxrule}
\end{equation}
so the replica takes over above the sharp threshold $\tc = \tM$, and the observable lifetime obeys $\tlife=\max\{\tM,\tau_m\}$.

Two questions classify the dynamics: is the dynamics memoryless, and which time sets the lifetime? The answers define three regimes,
\begin{equation}
\tlife=\left\{
\begin{array}{lll}
\tM, & \tau_m=0 & \text{(Markovian)},\\[2pt]
\tM, & 0<\tau_m\leq\tM & \text{(hidden memory)},\\[2pt]
\tau_m, & \tau_m>\tM & \text{(memory-dominated)}.
\end{array}\right.
\label{eq:regimes}
\end{equation}
In the \emph{Markovian} regime
the evolution reduces to the Markovian limit,
and the crystal decays at its Markovian lifetime.
In the \emph{hidden-memory} regime
the kernel has finite memory,
but the replica
decays at least as fast as
the Markovian subharmonic mode and
does not change the asymptotic exponential lifetime;
the measured $\tlife$ then provides only the upper bound
$\tau_m\le\tlife$.
In the \emph{memory-dominated} regime the replica
controls the asymptotic oscillatory decay,
and the kernel memory time
becomes the lifetime itself, $\tlife=\tau_m$.
The threshold $\tc=\tM$ thus marks where memory becomes
visible in the asymptotic decay rate, not where it appears:
this lifetime estimator cannot resolve memory times
shorter than $\tM$, although transient dynamics may reveal them.

The replica theorem constrains only the structure of
Eq.~\eqref{eq:embed}, not the Hamiltonian, the jump operators, or the Hilbert
space dimension. We therefore test it in two deliberately dissimilar DTCs: LMG Hamiltonian with incoherent pumping 
and a short-range tilted spin chain with dephasing (Sec.~\ref{sec:yousefjani}). Both are solved numerically at finite size without mean-field
or cumulant truncations, and the one-period propagator
$\mathfrak U_F(\tau_m)$ is diagonalized numerically.

\begin{figure}[t]
\centering
\includegraphics[width=\linewidth]{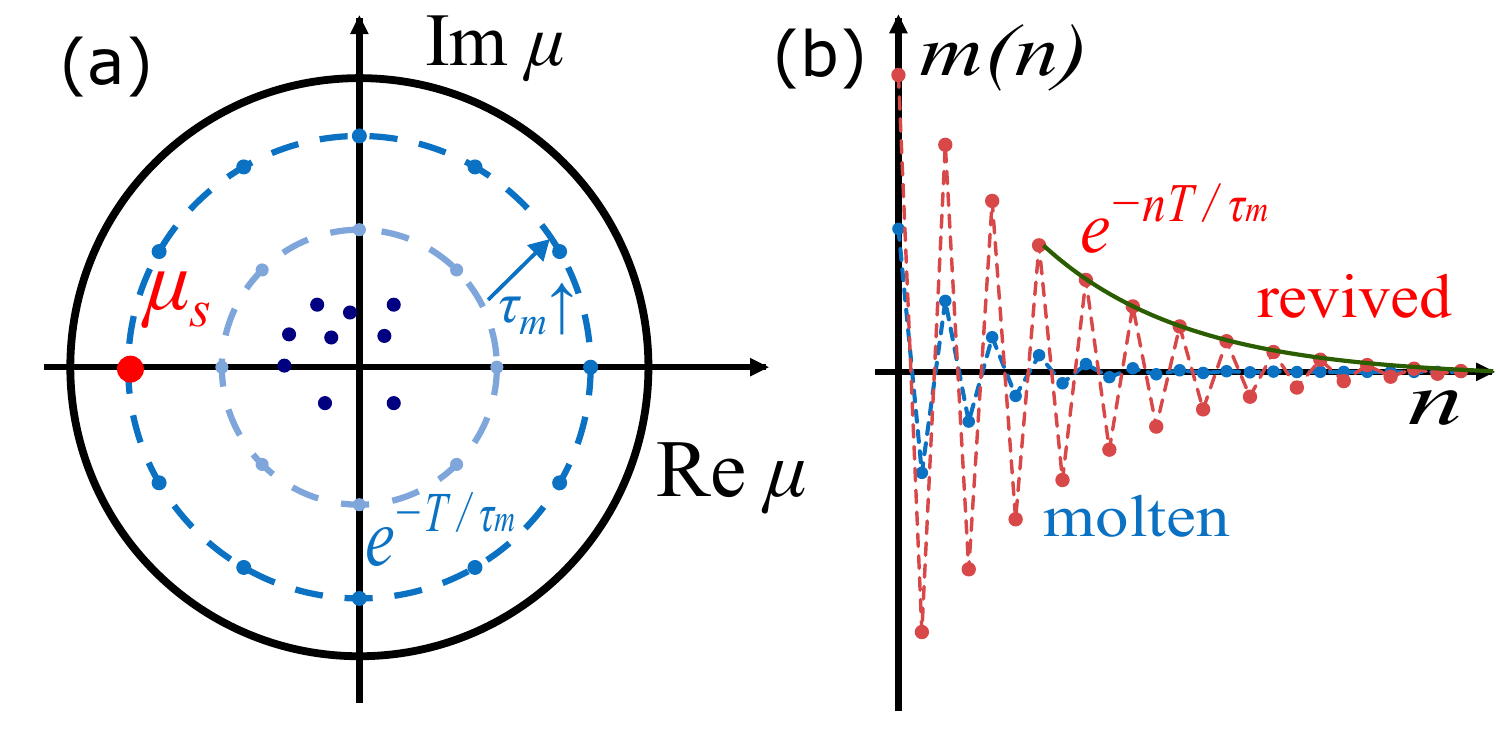}
\caption{\justifying{\textbf{The memory witness at a glance.}
\textbf{(a)}~By Eq.~\eqref{eq:replica} the Floquet spectrum of the
memory-embedded propagator is the Markovian cloud (grey) plus a replica of the
\emph{unitary} spectrum on a circle of radius $e^{-T/\tau_m}$ (blue). As
$\tau_m$ grows the replica circle inflates towards the unit circle (dashed);
once it exceeds the largest expressed oscillating Markovian modulus, the lifetime-setting mode
$\mu_s$ (red) locks onto it at angle $\pi$. \textbf{(b)}~The stroboscopic
response is period-doubled; its decay is set by $|\mu_s|$. In the hidden-memory regime ($\tau_m<\tc$) it
inherits the fast Markovian decay (grey, ``molten''), while in the memory-dominated regime $\tau_m>\tc$ the envelope follows $e^{-nT/\tau_m}$ (red, ``revived'') and reads
off the reservoir memory---independently of the drive and the dissipation. The
enabling step, an exact embedding of the post-Markovian dynamics into a
time-local generator coupling $\rho$ to an auxiliary memory sector $\eta$ at
rate $\tau_m^{-1}$, is described in the text.}}
\label{fig:schematic}
\end{figure}

\section{Models}
\subsection{Model~I: LMG with pumping}
Following Ref.~\cite{gong2018_discrete}, a robust dissipative time
crystal can arise when the interplay of interactions and incoherent
pumping stabilizes two $\mathbb Z_2$-symmetry-broken states
in the thermodynamic limit
and a parity operation toggles between them each period. Our first model realizes this with a collective spin
$J_\alpha=\tfrac12\sum_{j=1}^N\sigma_j^\alpha$  subject to the pump
$L_+=\sqrt{\gamma_+/S}\,J_+$ and the periodic drive
\begin{equation}
H(t)=-2h(t)\,J_z-\frac{\lambda}{S}J_x^2,
~
h(t)=
\begin{cases}
h_r, & 0\le t<t_r,\\
h_p, & t_r\le t<T,
\end{cases}
\label{eq:lmg_model}
\end{equation} 
with $T=t_r+t_p$ and $S=N/2$.  The longitudinal field $h(t)$ takes two values within one period: a weak residual field $h_r$ during the first segment, and a strong pulse field $h_p=\pi/(2t_p)$ during the second segment. In the strong-pulse limit $h_p\gg\lambda,\gamma_+$, with $h_pt_p=\pi/2$, the second segment approaches an ideal
$\pi$-rotation about $z$ (a parity kick) that swaps the two symmetry-broken states.
We monitor the stroboscopic magnetization $m_x(n)=\mathrm{Tr}[J_x\rho(nT)]/S$ from the $+x$ spin-coherent state.

\begin{figure}[t]
\centering
\includegraphics[width=\linewidth]{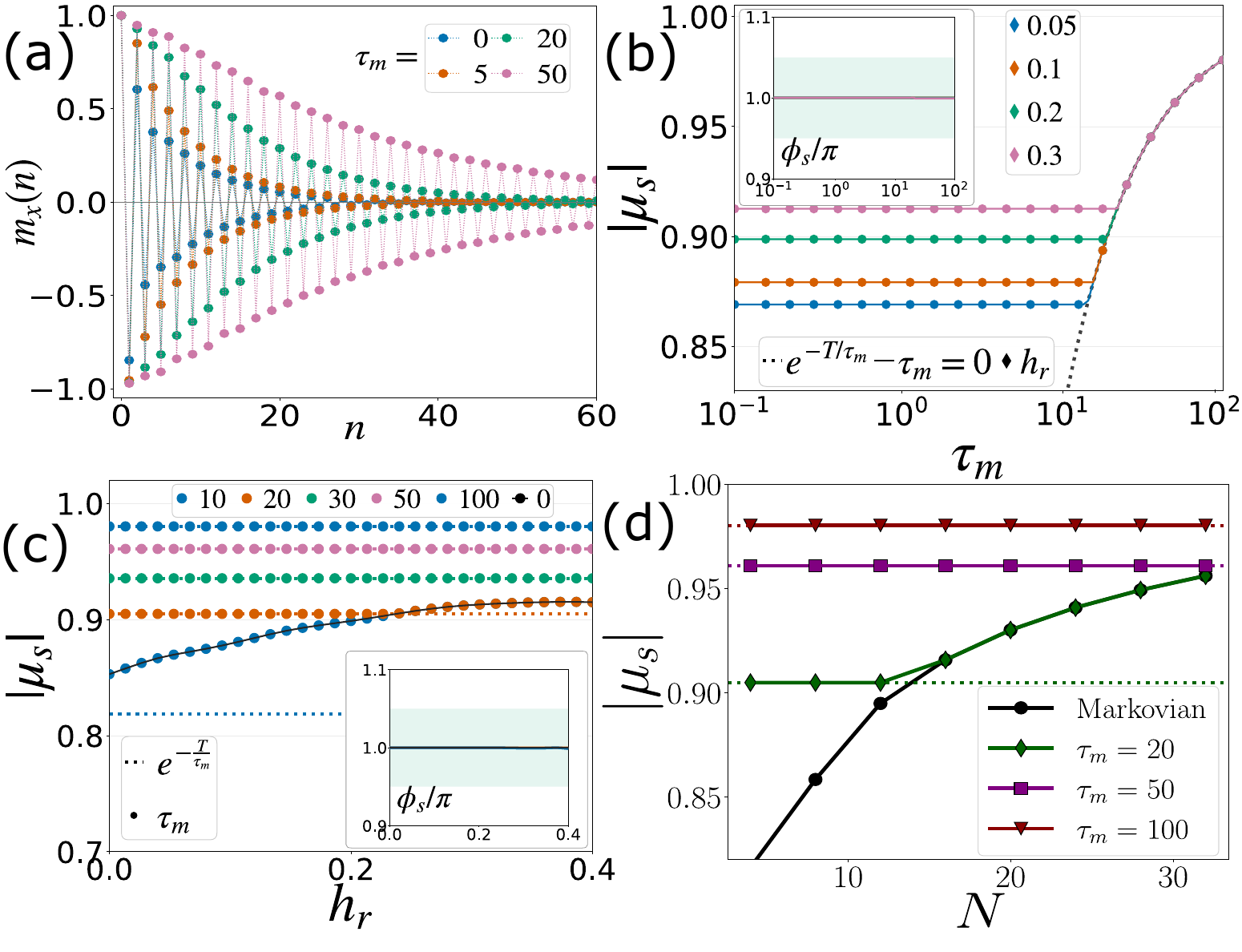}
\caption{\justifying{\textbf{Memory-revived time
crystal in the LMG model.}
\textbf{(a)}~Stroboscopic magnetization $m_x(n)$ for $\tau_m=0,5,20,50$ fixing $h_r=0.1$: the
Markovian response ($\tau_m=0$) melts within a few periods, whereas long memory
sustains a period-doubled signal whose envelope tracks $e^{-nT/\tau_m}$.
\textbf{(b)}~Floquet multiplier $|\mu_s|$ versus $\tau_m$ at fixed $h_r=0.05,0.1,0.2,0.3$ (colored diamonds): each
curve is flat at its Markovian value up to the kink at $\tc=\tM(h_r)$ and then locks
onto $e^{-T/\tau_m}$ (dotted).
\textbf{(c)}~$|\mu_s|$ versus the detuning $h_r$ for
$\tau_m=10,20,30,50,100$ (colored circles) and the Markovian reference
$\tau_m=0$ (solid); the memory branches sit on the parameter-free lines
$e^{-T/\tau_m}$ (dotted) while the Markovian branch drifts with the field.
\textbf{(d)}~$|\mu_s|$ versus system size $N$ fixing $h_r=0.1$: the Markovian modulus climbs as
the collective crystal stiffens toward its mean-field limit, whereas every
memory plateau stays pinned on $e^{-T/\tau_m}$, independent of $N$. Insets in
(b,c): subharmonic phase $\phi_s/\pi$, pinned to unity (shaded band: $\pm5\%$). All the common parameters are $N=10$, $t_r=t_p=1$, $\lambda=0.6$ and $\gamma_+=0.2$.}}
\label{fig:lmg}
\end{figure}

In Fig.~\ref{fig:lmg}, we show the
subharmonic dynamics and the underlying branch crossing.
In panel~(a) we plot the stroboscopic magnetization $m_x(n)$
as a function of $n$, for four memory times $\tau_m=0,5,20,50$, at fixed $h_r=0.1$. Under identical drive and dissipation
the Markovian evolution suppresses the subharmonic response
within a few periods, whereas $\tau_m=50$ sustains the subharmonic oscillations much longer. 
When the expressed replica modes dominate, the asymptotic exponential decay is governed by $e^{-nT/\tau_m}$, with lifetime $\tau_m$. The signal can additionally contain oscillatory modulations from the phases of these modes.

We note that in the Markovian limit the subharmonic lifetime
of this collective DTC grows with system size
[Fig.~\ref{fig:lmg}(d)], consistent with increasing stability of the finite-size subharmonic response. Establishing persistent thermodynamic
order and robustness against pulse errors requires further
analysis; a detailed analysis of the Markovian dynamics, including the
Floquet--Lindblad spectrum and finite-size scaling, will be presented separately~\cite{ghosh2026_DTC}. The spectral replica theorem and the finite-size memory-time estimator do not require a divergent
thermodynamic lifetime.

Panels~(b) and (c) resolve the mechanism. In panel~(b) we plot the modulus of the lifetime-setting Floquet multiplier, $|\mu_s|$, as a function of $\tau_m$, for $h_r=0.05,0.1,0.2,0.3$; the dotted line is the fit-free prediction $e^{-T/\tau_m}$ of Eq.~\eqref{eq:replica}.
Along the memory axis, $|\mu_s|$ remains pinned to its Markovian value up to the kink at $\tc=\tM(h_r)$ (hidden-memory regime), beyond which it follows $e^{-T/\tau_m}$ (memory-dominated regime). In panel~(c) we plot $|\mu_s|$ as a function of the magnetic
field $h_r$, for memory times  $\tau_m=10,20,30,50,100$, together with the Markovian reference $\tau_m=0$ (solid) and the corresponding levels $e^{-T/\tau_m}$ (dotted). 
Along the field axis, the Markovian branch drifts with $h_r$, a property of the driven system, while the replica-dominated segments form flat plateaus at $e^{-T/\tau_m}$, because at fixed $T$ and $\tau_m$
the drive enters the replica Floquet multiplier only through
its phase and never through its modulus. The exact spectral union permits a sharp crossing of the branch moduli, producing a kink in their maximum:
the kink occurs at $\tau_m=\tc$ in panel~(b), and at fields satisfying $\tM(h_r)=\tau_m$ in panel~(c) ~\cite{supplemental_mine2026}.
Equality of the moduli need not imply equality of the
complex multipliers, whose phases can differ. The insets of (b) and (c) show the phase $\phi_s/\pi$
of the same multiplier.

In panel~(d) we plot $|\mu_s|$ as a function of the system size $N$ at fixed $h_r=0.1$, for the Markovian case and for $\tau_m=20,50,100$. This separates the Markovian relaxation time from the kernel memory time.
The Markovian modulus climbs steadily with $N$
over the sizes studied,
so the threshold $\tc=\tM(N)$ increases with it.
The replica plateaus themselves remain fixed at
$e^{-T/\tau_m}$, independent of $N$,
because their modulus depends only on $T$ and $\tau_m$; their phases and  observable weights can still depend on $N$. The consequence is directly visible: the $\tau_m=20$ curve follows the replica plateau at small $N$
but switches to the Markovian branch once $\tM(N)$ exceeds $20$, i.e.\ the same kernel memory time passes from the memory-dominated to the hidden-memory regime as the crystal grows, while $\tau_m=50,100$ stay locked  to their replica plateaus across the whole window.
In the memory-dominated regime, the decay rate therefore reads out $\tau_m$, whereas the threshold is set by the Markovian dynamics; this inference requires no thermodynamic limit.
In the regime shown, a smaller crystal therefore has a lower memory-detection threshold: its shorter Markovian lifetime allows shorter
kernel memory times to be resolved through the asymptotic decay rate, provided the replica contribution remains measurable.

\subsection{Model~II: short-range chain with dephasing}
\label{sec:yousefjani}
For the second model, we consider
a finite-duration implementation of the disorder-free
Floquet DTC studied in Refs.~\cite{yousefjani2025_discrete,barlev2024_discrete}
and couple it to a dephasing environment.
In the unitary construction, a chain of $L$
spin-$\tfrac12$ particles evolves under a Stark-like Ising
interaction and a near-$\pi$ global spin rotation,
\begin{align}
H(t)&=
\begin{cases}
JH_I, & 0\le t<T/2,\\
H_P, & T/2\le t<T,
\end{cases}\nonumber\\
H_P&=(1-\varepsilon)\frac{\pi}{T}\sum_j\sigma_j^x,\quad
H_I=\sum_{j=1}^{L-1}j\,\sigma_j^z\sigma_{j+1}^z,
\label{eq:yousefjani_Hamiltonian}
\end{align}
where $\varepsilon$ is the global-pulse error and
the Ising couplings vary linearly along an open chain,
without quenched disorder.
We take $T=2$; the unitary Floquet operator is then
$U_F=e^{-iH_PT/2}e^{-i\Omega H_I}$ with $\Omega=JT/2$,
so that $\mathsf U_F^{\mathrm u}[\rho]=U_F\rho U_F^\dagger$.
This reproduces the unitary Floquet operator of the
kicked construction at the same $\Omega$ and $\varepsilon$, although the evolution within each period differs. At the resonant point $\Omega=\pi/2$ and for even $L$,
$(U_F)^2\propto\mathbb I$ for any $\varepsilon$,
giving exact two-period recurrence independently of
the pulse error ~\cite{yousefjani2025_discrete, barlev2024_discrete}. 
A nonzero period-doubled component additionally requires overlap of the initial state and observable with a $-1$ eigenmode of $\mathsf U_F^{\mathrm u}$.
Our numerics operate close to this point.

We now open this chain within the PMME framework of
Eq.~\eqref{eq:pmme}, coupling it to local pure-dephasing channels $L_j=\sqrt{\Gamma}\,\sigma_j^z$ with kernel memory time $\tau_m$. The dephasing acts during both steps of the drive, including the finite-duration spin flip. We monitor the stroboscopic longitudinal magnetization $m_z(n)=\frac1L\sum_j
\operatorname{Tr}\!\left[\sigma_j^z\rho(nT)\right]$ 
starting from the fully polarized state
$\rho(0)=
|\uparrow\cdots\uparrow\rangle
\langle\uparrow\cdots\uparrow|$.

\begin{figure}[t]
\centering
\includegraphics[width=\linewidth]{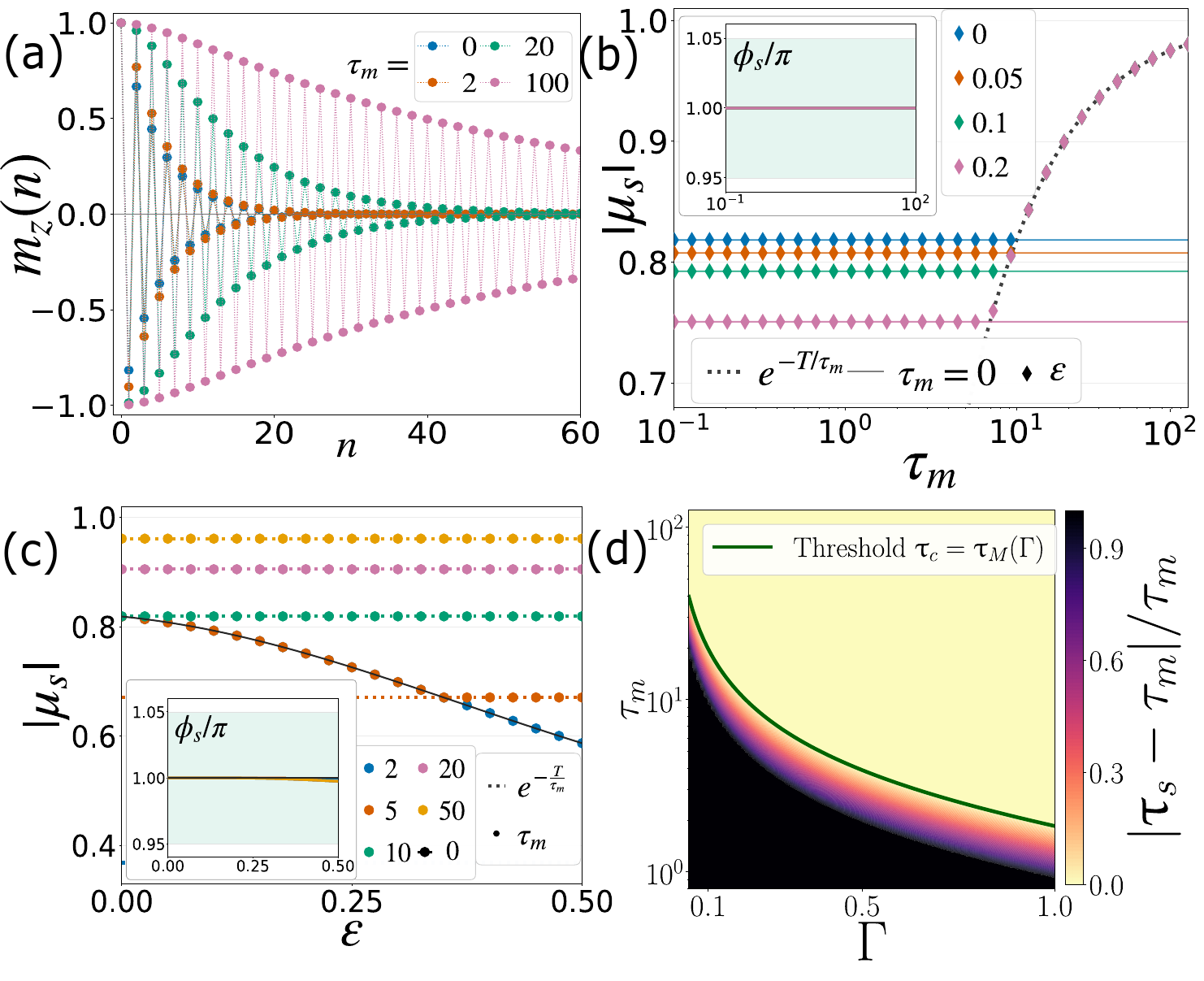}
\caption{\justifying{\textbf{Memory revived time crystal in a short-range chain with dephasing.}
\textbf{(a)}~Stroboscopic magnetization $m_z(n)$ for $\tau_m=0,2,20,100$ fixing $\varepsilon=0.01$; memory
revives the period-doubled response exactly as in the collective model.
\textbf{(b)}~Floquet multiplier $|\mu_s|$ versus $\tau_m$ at fixed $\varepsilon=0,0.05,0.1,0.2$ (colored diamonds): the same
plateau-then-$e^{-T/\tau_m}$ structure, with the kink at $\tc=\tM(\varepsilon)$.
\textbf{(c)}~$|\mu_s|$ versus the global-pulse error $\varepsilon$ for
$\tau_m=2,5,10,20,50$ (colored circles), the Markovian reference (solid), and
the predictions $e^{-T/\tau_m}$ (dotted); here the Markovian branch
\emph{decreases} with $\varepsilon$, the opposite trend to Fig.~\ref{fig:lmg}(c), yet
the memory plateaus land on identical values.
\textbf{(d)}~\emph{Regime boundary} in the $(\Gamma,\tau_m)$ plane fixing $\varepsilon=0.01$: the
threshold $\tc =\tM(\Gamma)$ (green) separates a hidden-memory regime 
(dark, large readout error $|\tlife-\tau_m|/\tau_m$) from a memory-dominated regime
(light, error $\to0$); the memory-dominated regime widens as dissipation grows.
Insets in (b,c): subharmonic phase $\phi_s/\pi$ (band: $\pm5\%$). 
All panels: $L=4$, $T=2$, $\Omega=\pi/2-0.01$, initial state $|\!\uparrow\cdots\uparrow\rangle$;
$\Gamma=0.2$. 
}}
\label{fig:sr}
\end{figure}

Figure~\ref{fig:sr} repeats the analysis of Fig.~\ref{fig:lmg} for a structurally
distinct DTC: nearest-neighbor instead of infinite-range coupling, local dephasing instead of collective pumping, a $2^L$-dimensional Hilbert
space, a different order parameter, and the pulse error $\varepsilon$ as the tuning
parameter of the Markovian lifetime. Since the replica theorem is exact for any such
choice, the nontrivial test here is that the replica branch is expressed in the
measured magnetization and sets its asymptotic exponential decay in the memory-dominated regime, exactly as in Model~I.

In panel~(a) we plot the stroboscopic longitudinal magnetization $m_z(n)$ as a function of $n$, for memory times $\tau_m=0,2,20,100$, at fixed pulse error $\varepsilon=0.01$. The $\tau_m=0,2$ traces decay substantially within $\sim15$ periods  ($\tau_m=2<\tM\simeq10$ lies in the hidden-memory regime and decays asymptotically at the Markovian rate), while $\tau_m=100$ persists across the plotted window. Here $\tM\simeq10$ corresponds to approximately five
drive periods, since $T=2$. In panel~(b) we plot $|\mu_s|$ as a function of $\tau_m$, for pulse errors $\varepsilon=0,0.05,0.1,0.2$, with the prediction $e^{- T/\tau_m}$ dotted. Each $\varepsilon$ holds a plateau at its own Markovian value up to the kink at $\tc=\tM(\varepsilon)$, beyond which \emph{all} curves collapse onto $e^{-T/\tau_m}$: in the memory-dominated regime the lifetime-setting modulus becomes independent of $\varepsilon$ and is fixed by $\tau_m$ at fixed $T$.

In panel~(c) we plot $|\mu_s|$ as a function of the pulse error $\varepsilon$, for memory times  $\tau_m=2,5,10,20,50$, together with
the Markovian reference (solid) and the levels $e^{-T/\tau_m}$ (dotted). The Markovian branch decreases with $\varepsilon$, while
the replica-dominated segments form plateaus at $e^{-T/\tau_m}$, following the same fit-free prediction as in Model~I. The threshold moves accordingly-- for $\tau_m=5$ the memory branch overtakes the Markovian one near $\varepsilon\simeq0.35$. 
A memory that is hidden in a more accurately driven crystal can become dominant as $\tM$ is reduced, provided the period-doubled response remains observable. The insets show that $\phi_s$ remains close to $\pi$ throughout the plotted DTC
regime: the change in lifetime is accompanied by an approximately period-doubled response. Thus two microscopically distinct DTCs with different Markovian responses recover the same memory-controlled modulus in the memory-dominated regime, allowing the kernel memory time to be inferred from
the asymptotic decay rate.

In panel~(d) we plot the relative
spectral inference
error $|\tlife-\tau_m|/\tau_m$ as a color map over the
$(\Gamma,\tau_m)$ plane at fixed $\varepsilon=0.01$,
with the threshold curve $\tc=\tM(\Gamma)$ superposed
(green), marking a sharp branch-crossing boundary
in the memory--dissipation plane.
Here $\tlife=-T/\ln|\mu_s|$ is obtained from the Floquet spectrum, rather than from a fit to a noisy time trace.
Below it we have the hidden-memory regime
[Eq.~\eqref{eq:regimes}], where the lifetime of the subharmonic response is determined by $\tM$; above it we have the memory-dominated regime, in which the
spectral inference error vanishes under the max rule,
up to numerical precision. The error also vanishes on the boundary,
where $\tM=\tau_m$. Because $\tM(\Gamma)$ decreases with $\Gamma$ in the regime shown, the boundary bends downward
and the memory-dominated region grows with $\Gamma$. Thus, in this regime, stronger dissipation lowers the
memory-detection threshold and enlarges the region in which the lifetime estimator operates. The memory-dominated
response therefore occupies an extended region of the
$(\Gamma,\tau_m)$ plane rather than an isolated parameter set.

\section{Discussion}

We have shown that a reservoir with a finite correlation time leaves an exact and universal fingerprint on the Floquet spectrum of a driven open time crystal. Embedding the master equation into a time-local generator on an enlarged operator space renders that generator exactly block-triangular, and the resulting Floquet spectrum is the Markovian spectrum plus a copy of the \emph{unitary} Floquet spectrum contracted by $e^{-T/\tau_m}$. Once the memory time exceeds the Markovian lifetime of the crystal (the memory-dominated regime), the replica branch controls the asymptotic subharmonic
decay, provided the relevant modes remain expressed in the observable. The resulting decay rate then measures the kernel memory time $\tau_m$. Because at fixed drive period $T$ the drive enters only the phase of the replica Floquet multiplier and the memory time only its modulus, the two pieces of information are carried by distinct spectral data.

The practical consequence is a witness that is unusually well matched to
what a time-crystal experiment already measures. The lifetime
$\tlife=-T/\ln|\mu_s|$ uses only the \emph{slope} of the period-doubled
decay and requires no tomography, no independent bath spectroscopy, and no detailed
Hamiltonian reconstruction once the applicability of the exponential-kernel model and replica
dominance have been established. Counterintuitively, stronger dissipation can lower the
memory-detection threshold when it shortens the competing Markovian lifetime, as demonstrated
in Model~II.  Existing trapped-ion, nitrogen-vacancy and superconducting-processor DTC platforms
~\cite{zhang2017_observation,choi2017_observation,
mi2022_time,frey2022_realization} measure stroboscopic spin observables, providing a natural setting in which to test this estimator.  Complementing these capabilities, Zhang et al.\
demonstrated that a phenomenological PMME accurately captures and predicts experimentally measured non-Markovian superconducting-qubit  dynamics~\cite{zhang2022_predicting}.
Vlachos et al.\ experimentally engineered non-Markovian dynamics using classical stochastic control of a superconducting transmon~\cite{vlachos2022_master}. Together, these developments motivate testing the proposed memory-time estimator in driven systems with controlled environmental noise.  The witness, however, operates only in the memory-dominated regime, $\tau_m>\tM$: the kernel memory time must exceed the corresponding Markovian lifetime. In the hidden-memory regime the measured asymptotic decay yields only the upper bound $\tau_m\le\tlife$.

Three caveats delimit the result. First, the replica branch must carry nonzero weight
in the measured observable. This weight is set by the modal coefficients and is  nonzero in the parameter regimes examined in both models studied here;
a period-doubled replica contribution requires
expressed coherent Floquet modes with phases at or near $\pi$. Neither their observable weight nor robust phase locking follows from the replica theorem alone. Second, the exact spectral splitting of Eq.~\eqref{eq:replica}, and the state decomposition
in Eq.~\eqref{eq:replica_state}, relies on the single-pole (exponential) memory kernel and the embedding used here. For kernels with several poles, or non-exponential kernels, this proof does not apply, and the decay need not equal the longest kernel time scale. Such extensions require a separate analysis.
Third, the lifetime signature identifies a kernel memory time within the assumed model; it does not by itself certify information backflow or CP indivisibility. These require independent diagnostics ~\cite{breuer2009_measure,rivas2014_quantum_non}.
The exponential-kernel construction also admits a
classical random-switching realization ~\cite{budini2014_post}, so the inferred time scale
does not identify the microscopic origin of the memory.

More broadly our results turn an environmental time scale into an observable feature of collective subharmonic response. Non-Markovianity is normally something to be diagnosed by
process-tomographic measures, such as trace-distance backflow~\cite{breuer2009_measure,liu2011_experimental}
or divisibility breaking~\cite{rivas2010_entanglement,hall2014_canonical}, that are
expensive to implement in a many-body setting~\cite{white2020_demonstration}. Here the
time crystal performs the measurement itself: its own rigidity converts the
bath correlation time into a directly readable decay rate. This suggests that
driven phases with a protected subharmonic response could serve as
built-in spectrometers of their own environment, and, conversely, that reservoir engineering is a route to extending the lifetime of the nonequilibrium order.

\appendix
\section{Exact time-local embedding and Floquet replica structure}
\subsection{Nonlocal equation and exact local embedding}

We begin from the modified post-Markovian equation
~\cite{shabani2005_completely,budini2014_post,sutherland2018_non,
zhang2022_predicting,nila2026_continuous}
\begin{align}
\dot{\rho}(t)
=
\mathcal{L}_H[\rho(t)]
+
\mathcal{L}_D\!\left[
\int_0^t dt'\,k(t-t')\,e^{\mathcal{L}(t-t')}\rho(t')
\right],
\label{SM:nonlocal_eq}
\end{align}
where  $\mathcal{L} = \mathcal{L}_H + \mathcal{L}_D, \mathcal{L}_H[\rho] = -i[H,\rho]$, and $\mathcal{L}_D[\rho] = \sum_i\left(
L_i\rho L_i^\dagger
-\frac{1}{2}\{L_i^\dagger L_i,\rho\}
\right)$. The memory kernel is chosen as $k(t)=\frac{1}{\tau_m}e^{-t/\tau_m}$, where $\tau_m$ is the reservoir correlation time. For this kernel it is convenient to define the operator $\eta(t)$, which we refer to as the pseudomode operator~\cite{budini2014_post},
\begin{align}
\eta(t)
=
\int_0^t dt'\,\frac{1}{\tau_m}
 e^{-(t-t')/\tau_m}
 e^{\mathcal{L}(t-t')}\rho(t'),
\label{SM:def_eta}
\end{align}
with initial condition $\eta(0)=0$. Differentiating
Eq.~\eqref{SM:def_eta} and using the Leibniz rule gives
\begin{align}
\dot{\eta}(t)
&=
\frac{1}{\tau_m}\rho(t)
+
\int_0^t dt'\,\frac{1}{\tau_m}
\left(\mathcal{L}-\tau_m^{-1}\mathbb{I}\right)
 e^{-(t-t')/\tau_m}e^{\mathcal{L}(t-t')}\rho(t')
\nonumber\\
&=
\tau_m^{-1}\rho(t)
+
\left(\mathcal{L}-\tau_m^{-1}\mathbb{I}\right)\eta(t).
\label{SM:eta_local}
\end{align}
Hence the nonlocal equation is exactly equivalent to
\begin{subequations} \label{SM:rho_eta_local}
\begin{align}
\dot{\rho}(t)
&=
\mathcal{L}_H[\rho(t)]+\mathcal{L}_D[\eta(t)],
\label{SM:rho_local}\\
\dot{\eta}(t)
&=
\tau_m^{-1}\rho(t)
+\left(\mathcal{L}-\tau_m^{-1}\mathbb{I}\right)\eta(t).
\label{SM:eta_local_final}
\end{align}
\end{subequations}

It is convenient to collect the physical state and pseudomode operator into
the embedded object $ \mathfrak P(t)$ and  write the dynamics in block form in doubled space
\begin{align}
\dot{\mathfrak P}(t)
=
\mathfrak L \,\mathfrak P(t),
\quad
\mathfrak P(t)
\equiv
\begin{pmatrix}
\rho(t)\\[2pt]
\eta(t)
\end{pmatrix},
\quad
\mathfrak L
=
\begin{pmatrix}
\mathcal{L}_H & \mathcal{L}_D\\[4pt]
\tau_m^{-1}\mathbb{I}
&
\mathcal{L}-\tau_m^{-1}\mathbb{I}
\end{pmatrix}.
\label{SM:block_generator}
\end{align}
The embedded generator $\mathfrak L$ acts on the doubled operator space $\mathfrak B\equiv
\mathcal{B}(\mathcal H)\oplus\mathcal{B}(\mathcal H)$. Equations~\eqref{SM:rho_eta_local}--\eqref{SM:block_generator} provide an exact time-local linear embedding of the original nonlocal dynamics, generated by $\mathfrak L$. The generator $\mathfrak L$ is not itself a GKLS generator on a doubled physical Hilbert space, since $\eta$ is a pseudomode operator and is generally not normalized.

\subsection{Trace preservation and complete positivity}

Trace preservation follows immediately from Eq.~\eqref{SM:rho_local},
\begin{align}
\frac{d}{dt}\operatorname{Tr}\rho(t)
=
\operatorname{Tr}\!\left[\mathcal{L}_H[\rho(t)]\right]
+
\operatorname{Tr}\!\left[\mathcal{L}_D[\eta(t)]\right]
=0,
\end{align}
because $\operatorname{Tr}\mathcal{L}_H[X]=0$ and
$\operatorname{Tr}\mathcal{L}_D[X]=0$ for any operator $X$.

The pseudomode operator is not normalized. Defining
$p(t)=\operatorname{Tr}\eta(t)$ and tracing
Eq.~\eqref{SM:eta_local_final} gives
\begin{align}
\dot p(t)=\frac{1-p(t)}{\tau_m},
\end{align}
and, with $p(0)=0$,
\begin{align}
p(t)=1-e^{-t/\tau_m}.
\label{SM:trace_eta}
\end{align}
Thus $\eta(t)$ has trace smaller than unity at finite time and approaches unit trace
asymptotically.

For the exponential-kernel construction used here, complete positivity can be
shown directly through a Markovian dilation. Define
\begin{align}
\varsigma(t)\equiv\rho(t)-\eta(t).
\label{SM:def_varsigma}
\end{align}
Subtracting Eq.~\eqref{SM:eta_local_final} from
Eq.~\eqref{SM:rho_local} gives
\begin{subequations} \label{SM:varsigma_eta_triangular}
\begin{align}
\dot\varsigma(t)
&=
\left(\mathcal L_H-\tau_m^{-1}\mathbb I\right)\varsigma(t),
\label{SM:varsigma_local}\\
\dot\eta(t)
&=
\tau_m^{-1}\varsigma(t)+\mathcal L\eta(t).
\label{SM:eta_triangular}
\end{align}
\end{subequations}
Introduce a two-state ancilla and denote its matrix units by
$E_{ab}$, with $a,b\in\{0,1\}$. The block-diagonal state is
\begin{align}
\varrho(t)
=
\varsigma(t)\otimes E_{00}
+
\eta(t)\otimes E_{11}.
\label{SM:dilation_state}
\end{align}
With
\begin{align}
H_{\rm dil}&=H\otimes\mathbb I_a,\\
J_m&=\tau_m^{-1/2}\mathbb I\otimes E_{10},\\
J_i&=L_i\otimes E_{11},
\end{align}
the GKLS equation for $\varrho(t)$ reproduces
Eqs.~\eqref{SM:varsigma_local}--\eqref{SM:eta_triangular} for the two diagonal
blocks and leaves the block-diagonal form \eqref{SM:dilation_state} invariant.
Because $\eta(0)=0$, the initial dilated state is the fixed product
\begin{align}
\varrho(0)=\rho_0\otimes E_{00}\ \ge 0,
\label{SM:dilation_init}
\end{align}
i.e.\ the system in $\rho_0$ and the ancilla in the fixed pure state $E_{00}$. The generator $(H_{\rm dil},J_m,J_i)$ is of GKLS form, so $\varrho(t)$ evolves completely positively; in particular $\varrho(t)\ge0$ and each diagonal block is positive, whence $\varsigma(t)\ge0$ and $\eta(t)\ge0$ separately. Thus $\eta(t)\ge0$ with $\operatorname{Tr}\eta(t)\le1$, so it may
consistently be regarded as a subnormalized pseudomode operator. With the ancilla prepared in a state independent of $\rho_0$ and
\begin{align}
\rho(t)=\operatorname{Tr}_a\varrho(t)=\varsigma(t)+\eta(t),
\end{align}
the map $\rho_0\mapsto\rho(t)$ is a fixed-dilation followed by a partial trace,
and is therefore CPTP. The construction carries over verbatim to a driven
Hamiltonian, with $H_{\rm dil}(t)=H(t)\otimes\mathbb I_a$.

\subsection{Embedded spectral problem and exact branch decomposition}

In any fixed basis of $\mathfrak B$, let
$\mathfrak P_\alpha^{\mathrm R}$ and
$\mathfrak Q_\alpha^{\mathrm L}$
denote, respectively, the right eigenmode of $\mathfrak L$ and its left dual:
\begin{align}
\mathfrak L\,\mathfrak P_\alpha^{\mathrm R}
&=
\Lambda_\alpha\mathfrak P_\alpha^{\mathrm R},
&
\mathfrak Q_\alpha^{\mathrm L}
\mathfrak L
&=
\Lambda_\alpha
\mathfrak Q_\alpha^{\mathrm L}.
\label{SM:embedded_eigenproblem}
\end{align}
For a diagonalizable $\mathfrak L$ we choose the biorthogonal normalization
\begin{align}
\mathfrak Q_\alpha^{\mathrm L}
\mathfrak P_\beta^{\mathrm R}
=
\delta_{\alpha\beta}.
\label{SM:embedded_biorthogonality}
\end{align}
The right mode is written as
\begin{align}
\mathfrak P_\alpha^{\mathrm R}
=
\begin{pmatrix}
\sigma_\alpha^{\mathrm R}\\[2pt]
\zeta_\alpha^{\mathrm R}
\end{pmatrix},
\end{align}
where $\sigma_\alpha^{\mathrm R}$ and $\zeta_\alpha^{\mathrm R}$ are the
physical and pseudomode components. The right-eigenvalue equations are
\begin{subequations}
\begin{align}
\Lambda_\alpha\sigma_\alpha^{\mathrm R}
&=
\mathcal L_H\sigma_\alpha^{\mathrm R}
+\mathcal L_D\zeta_\alpha^{\mathrm R},
\label{SM:eig1}\\
\Lambda_\alpha\zeta_\alpha^{\mathrm R}
&=
\tau_m^{-1}\sigma_\alpha^{\mathrm R}
+\left(\mathcal L-\tau_m^{-1}\mathbb I\right)
\zeta_\alpha^{\mathrm R}.
\label{SM:eig2}
\end{align}
\end{subequations}
Whenever the shifted resolvent exists,
Eq.~\eqref{SM:eig2} gives
\begin{align}
\zeta_\alpha^{\mathrm R}
=
\tau_m^{-1}
\left(
\Lambda_\alpha\mathbb I-\mathcal L+\tau_m^{-1}\mathbb I
\right)^{-1}
\sigma_\alpha^{\mathrm R}.
\label{SM:zeta_resolvent}
\end{align}
Substitution into Eq.~\eqref{SM:eig1} yields the reduced nonlinear equation
\begin{align}
\left[
\Lambda\mathbb I
-\mathcal L_H
-\tau_m^{-1}\mathcal L_D
\left(
\Lambda\mathbb I-\mathcal L+\tau_m^{-1}\mathbb I
\right)^{-1}
\right]\sigma^{\mathrm R}=0.
\label{SM:nonlinear_branch}
\end{align}

The same elimination can be expressed through the
Schur-complement determinant identity,
\begin{align}
\det\!\left(\Lambda\mathbb I-\mathfrak L\right)
&=
\det\!\left(
\Lambda\mathbb I-\mathcal L+\tau_m^{-1}\mathbb I
\right)
\times
\\ \nonumber
&\det\!\left[
\Lambda\mathbb I-\mathcal L_H
-\tau_m^{-1}\mathcal L_D
\left(
\Lambda\mathbb I-\mathcal L+\tau_m^{-1}\mathbb I
\right)^{-1}
\right],
\label{SM:full_char_eq}
\end{align}
whenever the Schur complement is well defined. This is a meromorphic
representation of the characteristic determinant: zeros and poles of the two
factors must not be interpreted separately at singular points of the
resolvent.

The exact spectral structure is more transparent after the change of variables
$\varsigma=\rho-\eta$. Define
\begin{align}
\mathfrak P_{\triangle}(t)
\equiv
\begin{pmatrix}
\varsigma(t)\\[2pt]
\eta(t)
\end{pmatrix}
=
\mathfrak S\,\mathfrak P(t),
\qquad
\mathfrak S
=
\begin{pmatrix}
\mathbb I&-\mathbb I\\
0&\mathbb I
\end{pmatrix}.
\end{align}
Then
\begin{align}
\dot{\mathfrak P}_{\triangle}(t)
=
\mathfrak L_{\triangle}\mathfrak P_{\triangle}(t),
\qquad
\mathfrak L_{\triangle}
=
\mathfrak S\mathfrak L\mathfrak S^{-1}
=
\begin{pmatrix}
\mathcal L_H-\tau_m^{-1}\mathbb I&0\\[4pt]
\tau_m^{-1}\mathbb I&\mathcal L
\end{pmatrix}.
\label{SM:triangular_generator}
\end{align}
Because $\mathfrak L_{\triangle}$ is block lower triangular,
\begin{align}
\det\!\left(\Lambda\mathbb I-\mathfrak L\right)
=
\det\!\left(
\Lambda\mathbb I-\mathcal L_H+\tau_m^{-1}\mathbb I
\right)
\det\!\left(\Lambda\mathbb I-\mathcal L\right),
\label{SM:exact_char_factor}
\end{align}
and therefore
\begin{align}
\operatorname{spec}\mathfrak L
=
\operatorname{spec}\mathcal L
\;\cup\;
\operatorname{spec}\!\left(
\mathcal L_H-\tau_m^{-1}\mathbb I
\right).
\label{SM:static_spectrum}
\end{align}
This factorization is exact and requires no commutativity assumption between
$\mathcal L_H$ and $\mathcal L_D$. The Schur-complement form
Eq.~\eqref{SM:full_char_eq} and the triangular factorization
Eq.~\eqref{SM:exact_char_factor} are two representations of the same
characteristic determinant; the latter makes the two exact spectral families
manifest.

\subsection{Markovian limit and stationary states}

The Markovian limit is obtained by taking $\tau_m\to0$, for which $k(t)=\frac{1}{\tau_m}e^{-t/\tau_m} \longrightarrow\delta(t)$ as a one-sided approximate identity. The triangular variables make this limit particularly transparent.  Equation~\eqref{SM:varsigma_local}
has the exact solution
\begin{align}
\varsigma(t)
=
e^{-t/\tau_m}e^{t \mathcal L_H}\rho_0,
\qquad
\varsigma(0)=\rho_0.
\label{SM:varsigma_solution}
\end{align}
Using $\eta=\rho-\varsigma$, Eq.~\eqref{SM:rho_local} becomes $
\dot\rho(t) = \mathcal L\rho(t) - \mathcal L_D\varsigma(t)$. For every fixed $t>0$, $\varsigma(t)\to0$ as $\tau_m\to0$, while its
integrated contribution is $\mathcal O(\tau_m)$ in finite dimension.
Consequently
\begin{align}
\rho(t)\longrightarrow\rho_M(t),
\qquad
\dot\rho_M(t)=\mathcal L\rho_M(t),
\label{SM:markov_limit}
\end{align}
for fixed finite times. Equation~\eqref{SM:trace_eta} reflects the same initial boundary layer of width $\mathcal O(\tau_m)$.

For finite $\tau_m$, the stationary subspace is inherited exactly  from the
Markovian generator. From Eqs.~\eqref{SM:varsigma_eta_triangular}, a  stationary embedded state satisfies
\begin{align}
0
=
\left(\mathcal L_H-\tau_m^{-1}\mathbb I\right)\varsigma_\infty,
\qquad
0
=
\tau_m^{-1}\varsigma_\infty+\mathcal L\eta_\infty.
\end{align}
Since $\mathcal L_H$ has purely imaginary spectrum,
$\mathcal L_H-\tau_m^{-1}\mathbb I$ has no zero eigenvalue for finite
$\tau_m$. Hence
\begin{align}
\varsigma_\infty=0,
\qquad
\mathcal L\eta_\infty=0,
\qquad
\rho_\infty=\eta_\infty.
\label{SM:steady_states}
\end{align}
Thus every stationary state of the embedded dynamics is a stationary state of
$\mathcal L$, and conversely every Markovian stationary state lifts to the
embedded stationary pair
\begin{align}
\mathfrak P_\infty
=
\begin{pmatrix}
\rho_\infty\\[2pt]
\rho_\infty
\end{pmatrix}.
\end{align}

\subsection{Driven extension, Floquet  spectrum, and expressed modes} 

The derivation above was formulated for a time-independent $\mathcal L$. For an
explicitly driven Hamiltonian, the ordinary propagator is replaced by the
two-time propagator
\begin{align}
\Phi(t,t')
\equiv
\mathcal T\exp\!\left[
\int_{t'}^t ds\,\mathcal L(s)
\right],
\qquad
\mathcal L(t)=\mathcal L_H(t)+\mathcal L_D.
\label{SM:time_dep_prop}
\end{align}
The driven extension of Eq.~\eqref{SM:nonlocal_eq} is
\begin{align}
\dot\rho(t)
=
\mathcal L_H(t)[\rho(t)]
+
\mathcal L_D\!\left[
\int_0^t dt'\,k(t-t')\Phi(t,t')\rho(t')
\right].
\label{SM:nonlocal_eq_td}
\end{align}
For the same exponential kernel, define the pseudomode operator
\begin{align}
\eta(t)
=
\int_0^t dt'\,\frac{1}{\tau_m}
 e^{-(t-t')/\tau_m}\Phi(t,t')\rho(t').
\label{SM:def_eta_td}
\end{align} 
The exact local  equations retain the same  structure,
\begin{subequations}
\begin{align}
\dot\rho(t)
&=
\mathcal L_H(t)[\rho(t)]+\mathcal L_D[\eta(t)],
\label{SM:rho_local_td}\\
\dot\eta(t)
&=
\tau_m^{-1}\rho(t)
+\left(\mathcal L(t)-\tau_m^{-1}\mathbb I\right)\eta(t).
\label{SM:eta_local_td}
\end{align}
\end{subequations}
Accordingly,
\begin{align}
\dot{\mathfrak P}(t)
=
\mathfrak L(t)\mathfrak P(t),
\qquad
\mathfrak L(t)
=
\begin{pmatrix}
\mathcal L_H(t)&\mathcal L_D\\[4pt]
\tau_m^{-1}\mathbb I&\mathcal L(t)-\tau_m^{-1}\mathbb I
\end{pmatrix}.
\label{SM:block_generator_td}
\end{align}
The same time-independent similarity transformation $\mathfrak S$  introduced
above triangularizes $\mathfrak L(t)$ at every instant,
\begin{align}
\mathfrak S\mathfrak L(t)\mathfrak S^{-1}
=
\begin{pmatrix}
\mathcal L_H(t)-\tau_m^{-1}\mathbb I&0\\[4pt]
\tau_m^{-1}\mathbb I&\mathcal L(t)
\end{pmatrix}.
\label{SM:triangular_generator_td}
\end{align}

For a periodic drive, $H(t+T)=H(t)$, define the embedded one-period propagator
\begin{align}
\mathfrak U_F(\tau_m)
=
\mathcal T\exp\!\left[
\int_0^Tdt\,\mathfrak L(t)
\right],
\label{SM:floquet_prop}
\end{align}
and the purely unitary and Markovian Floquet channels
\begin{align}
\mathsf U_F^{\mathrm u}
&=
\mathcal T\exp\!\left[
\int_0^Tdt\,\mathcal L_H(t)
\right],
&
\mathsf U_F^{M}
&=
\mathcal T\exp\!\left[
\int_0^Tdt\,\mathcal L(t)
\right].
\label{SM:sector_floquet_ops}
\end{align} 
Because the same transformation $\mathfrak S$ applies at all times,
  \begin{align}
\mathfrak S\mathfrak U_F(\tau_m)\mathfrak S^{-1}
=
\begin{pmatrix}
 e^{-T/\tau_m}\mathsf U_F^{\mathrm u}&0\\[4pt]
 \mathsf C_F&\mathsf U_F^{M}
\end{pmatrix},
\label{SM:floquet_triangular}
\end{align} 
where $\mathsf C_F$ is an off-diagonal feed-forward block and therefore does
not affect the spectrum. Hence
\begin{align}
\operatorname{spec}\mathfrak U_F(\tau_m)
=
\operatorname{spec}\mathsf U_F^{M}
\;\cup\;
e^{-T/\tau_m}\operatorname{spec}\mathsf U_F^{\mathrm u}.
\label{SM:replica_spectrum}
\end{align}
Since $\mathsf U_F^{\mathrm u}$ is a unitary conjugation channel, every replica multiplier
has modulus $|\mu^{\mathrm{rep}}|=e^{-T/\tau_m}$. 
 Moreover, $\eta(0)=0$ implies $\varsigma(0)=\rho_0$, and therefore
\begin{align}
\rho(nT)
=
\underbrace{
e^{-nT/\tau_m}
\left(\mathsf U_F^{\mathrm u}\right)^n[\rho_0]
}_{\displaystyle \varsigma(nT)}
+\eta(nT).
\label{SM:replica_state}
\end{align} 
This identity is exact: the  first term is the autonomous replica contribution, whose envelope is fixed solely by $\tau_m$.

\paragraph{Floquet-mode expansion.}
Away from isolated defective degeneracies, we assume that $\mathfrak U_F$ is diagonalizable. Let
$\mathfrak P_{\alpha,F}^{\mathrm R}$ denote its right column modes and $\mathfrak Q_{\alpha,F}^{\mathrm L}$ the corresponding left dual modes,
\begin{align}
\mathfrak U_F\mathfrak P_{\alpha,F}^{\mathrm R}
&=
\mu_\alpha\mathfrak P_{\alpha,F}^{\mathrm R},
&
\mathfrak Q_{\alpha,F}^{\mathrm L}\mathfrak U_F
&=
\mu_\alpha\mathfrak Q_{\alpha,F}^{\mathrm L},
\label{SM:floquet_eigenproblem}
\end{align} 
with biorthogonal normalization 
$\mathfrak Q_{\alpha,F}^{\mathrm L}
\mathfrak P_{\beta,F}^{\mathrm R}
=\delta_{\alpha\beta}.
$
Write the physical and pseudomode components of  a right Floquet mode as
  \begin{align}
\mathfrak P_{\alpha,F}^{\mathrm R}
=
\begin{pmatrix}
\sigma_{\alpha,F}^{\mathrm R}\\[2pt]
\zeta_{\alpha,F}^{\mathrm R}
\end{pmatrix}.
\label{SM:floquet_mode_components}
\end{align} 
For the initially empty pseudomode,
\begin{align}
\mathfrak P(0)
=
\begin{pmatrix}
\rho_0\\[2pt]
0
\end{pmatrix},
\end{align} 
and its modal expansion is
$
\mathfrak P(0)
=
\sum_\alpha a_\alpha\mathfrak P_{\alpha,F}^{\mathrm R},
 a_\alpha
=
\mathfrak Q_{\alpha,F}^{\mathrm L}\mathfrak P(0).
$
For an  observable $O$ acting on the physical sector,
\begin{align}
m_O(n)
\equiv
\operatorname{Tr}\!\left[O\rho(nT)\right]
=
\sum_\alpha c_\alpha^{(O)}\mu_\alpha^n,
\label{SM:observable_spectral}
\end{align}
where 
$
c_\alpha^{(O)}
=
a_\alpha\,
\operatorname{Tr}\!\left[
O\sigma_{\alpha,F}^{\mathrm R}
\right].
$
Thus $c_\alpha^{(O)}$ is the weight with which Floquet mode $\alpha$ is
expressed in the measured signal; a mode contributes only when
$c_\alpha^{(O)}\neq0$. For exactly degenerate, or numerically unresolved,
multipliers, individual eigenvectors within the degenerate subspace are not
unique. We therefore group such multipliers and sum their complex
contributions before assigning an observable weight, making the result
independent of the basis chosen within that subspace. The numerical threshold
used to declare a mode expressed is specified below in
Eq.~\eqref{SM:expressed_set}. Writing
\begin{align}
\mu_\alpha=|\mu_\alpha|e^{i\phi_\alpha},
\quad
\phi_\alpha=\arg\mu_\alpha\in(-\pi,\pi],
\quad
\kappa_\alpha=-\ln|\mu_\alpha|,
\quad
 g_\alpha=\frac{\kappa_\alpha}{T},
\label{SM:mode_decay_phase}
\end{align}
separates the decay rate $g_\alpha$ per unit time from the oscillation phase
$\phi_\alpha$. In particular, $|\phi_\alpha|=\pi$ produces the period-doubled
factor $(-1)^n$.

\paragraph{Mode selection.}
We first define the set of dynamically expressed, nonstationary modes,
\begin{align}
\mathcal E_O
=
\left\{
\alpha:
|c_\alpha^{(O)}|
\ge
\epsilon_{\rm w}\sum_\beta|c_\beta^{(O)}|
\quad\text{and}\quad
|\mu_\alpha|<1-\delta_{\rm s}
\right\}.
\label{SM:expressed_set}
\end{align}
Here $\epsilon_{\rm w}$ and $\delta_{\rm s}$ are, respectively, the observable-weight and stationary-mode cutoffs used in the numerical analysis.

To identify the oscillatory sector relevant to the DTC response, we further define
\begin{align}
\mathcal E_O^{\rm osc}
=
\left\{
\alpha\in\mathcal E_O:
|\phi_\alpha|>\phi_{\rm min}
\right\},
\label{SM:oscillating_set}
\end{align}
where $\phi_{\rm min}$ excludes modes whose phase is numerically indistinguishable from zero; the model-specific numerical tolerances are given with the corresponding calculations below. For Model~I we take $O=S_x/S$, the normalized transverse magnetization, with $S=N/2$. For the finite-size parity-symmetric dynamics considered here, the stationary mode is even and therefore has zero overlap with this odd observable. The same construction is applied to the longitudinal DTC order parameter of Model~II.

The lifetime-setting multiplier is the slowest expressed oscillating mode,
\begin{align}
\mu_s=\mu_{\alpha_s},\qquad
\alpha_s=\operatorname*{arg\,max}_{\alpha\in\mathcal E_O^{\rm osc}}|\mu_\alpha| .
\label{SM:muslow}
\end{align}
When several multipliers share this modulus within a relative tolerance $10^{-6}$ (as for the
replica branch, all of whose multipliers have modulus $e^{-T/\tau_m}$), we select the one with
the largest $|c_\alpha^{(O)}|$; its phase defines $\phi_s$.

\paragraph{Lifetime-setting branch and activation threshold.}
Within the DTC regime considered in the main text, let $\mu_M$ denote the slowest expressed period-doubled multiplier of the purely Markovian Floquet channel $\mathsf U_F^{M}$. Define the Markovian DTC lifetime
\begin{align}
\tM
=
-\frac{T}{\ln|\mu_M|}.
\label{SM:tauM}
\end{align}
Provided the relevant Markovian and replica period-doubled modes are both expressed for the chosen initial-state--observable pair, and remain the lifetime-setting oscillatory candidates in the parameter regime under consideration, the two exact spectral families compete only through their moduli. The lifetime-setting DTC multiplier therefore obeys $|\mu_s(\tau_m)|
=
\max\!\left\{
|\mu_M|,
e^{-T/\tau_m}
\right\}$. Consequently,
\begin{align}
\tlife
\equiv
-\frac{T}{\ln|\mu_s|}
=
\max\{\tM,\tau_m\},
\label{SM:lifetime_rule}
\end{align}
The activation threshold is $\tc=\tM$. Below threshold (the \emph{hidden-memory regime} of the main text, $0<\tau_m<\tM$) the measured DTC lifetime is pinned to $\tM$ and provides only the upper bound $\tau_m\le\tlife$; above threshold (the \emph{memory-dominated regime}) the replica branch sets the decay and $\tlife=\tau_m$. At an isolated defective degeneracy, generalized Floquet modes replace the simple modal expansion, but the exact spectral radii and the threshold condition above are unchanged.

\section*{Acknowledgment}
This research was supported by the Israel Science Foundation (grant No. 2120/24). RG warmly thank Subhajit Sarkar for his valuable insights. V.M. acknowledges support from Anusandhan National Research Foundation (ANRF) through ARG (Project No. ANRF/ARG/2025/002531/PS).

\bibliographystyle{apsrev4-2}
\bibliography{refs_non-Mar}
\end{document}